\documentclass[12pt]{article}
\usepackage{a4wide,epsfig,amsmath,amssymb,cite,scalefnt,graphicx}

\def\npb{{\it{Nucl.\ Phys.}\ }{\bf B}}
\def\plb{{{\it Phys.\ Lett.}\ }{ \bf B}}
\def\jhep{{\it JHEP}\ }
\def\be{\begin{equation}}
\def\ee{\end{equation}}
\def\bea{\begin{eqnarray}}
\def\eea{\end{eqnarray}}
\def\nn{\nonumber\\}

\def\tr{\hbox{tr}}
\def\barh{\overline h} 
\def\prd{{{\it Phys.\ Rev.}\ }{\bf D}}

\def\thetabar{{\overline\theta}}

\def\cbar{\overline c}
\def\fbar{\overline f}
\def\Fbar{\overline F}
\def\Lambdabar{\overline \Lambda}

\def\Tr{\mathop{\rm Tr}}

\def\Bbar{\overline B}

\def\fhat{\hat f}

\def\frakk[#1#2{{{#1}\over{#2}}}

\def\lambdabar{\overline\lambda}

\def\Dbar{\overline D}

\def\Fbar{\overline F}

\def\Ncal{{\cal N}}

\def\Vhat{\hat V}
\def\Rhat{\hat R}
\def\Abar{\overline A}
\def\Bbar{\overline B}
\def\Mbar{\overline M}
\def\Nbar{\overline N}
\def\Hbar{\overline H}
\def\ahat{\hat a}
\def\bhat{\hat b}

\def\pa{\partial}
\def\pabar{\overline\partial}

\input epsf
\begin{document}

\begin{titlepage}
\begin{flushright}
LTH987\\
\end{flushright}
\date{}
\vspace*{3mm}

\begin{center}
{\Huge Superconformal Chern-Simons theories beyond
leading order}\\[12mm]
{\bf I.~Jack\footnote{{\tt dij@liv.ac.uk}} and 
C.~Luckhurst\footnote{{\tt mf0u60d7@liv.ac.uk}}}\\

\vspace{5mm}
Dept. of Mathematical Sciences,
University of Liverpool, Liverpool L69 3BX, UK\\

\end{center}

\vspace{3mm}
\begin{abstract}
We discuss higher-order corrections to superconformal invariance for
a class of $\Ncal=2$ supersymmetric Chern-Simons theories including
the ABJM model. We argue that corrections are inevitable for general theories
in this class; but that it is probable that any corrections
are of a particular ``maximally transcendental'' form. 

\end{abstract}

\vfill

\end{titlepage}

\section{Introduction}
Chern-Simons gauge theories have attracted attention for a considerable 
time due to their topological nature\cite{schwarza, witten, djt}
(in the pure gauge case) and their 
possible relation to the quantum Hall effect and high-$T_c$ 
superconductivity. More recently  
there has been substantial interest in $\Ncal=2$ supersymmetric Chern-Simons
matter theories in the context of the AdS/CFT 
correspondence and in particular, a wide range of superconformal theories has 
been discovered\cite{schwarz}-\cite{MFMa}, starting with the 
BLG\cite{bagger,gust} and ABJ/ABJM\cite{ABJ,ABJM} models.
Although a more familiar formulation is in terms of ``quiver''-type gauge 
theories based on the gauge group $U(N)\times U(M)$, many of them may be 
understood in terms of an underlying ``3-algebra'' 
structure\cite{bagger},\cite{baggera}-\cite{CWf}. 
It was shown in Ref.~\cite{GY} that $\Ncal=3$ Chern-Simons theories
(which include the ABJ, ABJM models as special cases) are 
are exactly superconformal to all orders. 
Explicit perturbative computations to
corroborate the superconformal property have been carried out in 
Refs.~\cite{ASW, penatia, penatib}
at lowest order (two loops for a theory in three dimensions). Since
the gauge coupling $\beta$-function is zero for any Chern-Simons 
theory\cite{Kap} due to
the topological nature of the theory (and indeed is quantised at certain 
values--the Chern-Simons ``level'') it is only necessary to compute 
the anomalous dimensions of the chiral fields in order to check for 
superconformality (in view of the non-renormalisation theorem). Our purpose
here is to attempt to extend the explicit check of superconformality
beyond lowest order, and beyond the $\Ncal=3$ theories which are already known
to be exactly superconformal. 
  These $\Ncal=3$ superconformal theories involve a simple
choice of the superpotential couplings in terms of the Chern-Simons level, and
the fact that this choice renders the theory finite to all orders is
analogous to the case 
of $\Ncal=4$ and $\Ncal=2$ 
supersymmetric theories in four dimensions, where the finiteness properties 
are manifest to all orders in the $\Ncal =1$
superfield description once the field content and superpotential
have been specified (assuming 
a supersymmetric regulator such as DRED). However, an alternative 
possibility is that one might have to adjust the couplings order by order so 
as to achieve finiteness\cite{DRTJ,EKT}. 
This would be more analogous to the case of finite $\Ncal =1$ theories
in four dimensions, where the finiteness
is obtained through an order-by-order adjustment of the couplings. We might 
well expect this behaviour in theories where superconformality is 
achieved by solving a somewhat non-trivial condition at lowest order. 

In odd spacetime dimensions, divergences only occur at even loop order, so 
to go beyond leading order we are driven to consider a four-loop 
calculation. The total number of diagrams is colossal; so here we report on
what can be learned from the consideration of a subset of the full set of 
diagrams, namely those which have at least one (in fact at least two) Yukawa
vertices. We were able to compute all the relevant diagrams with the exception
of a single non-planar diagram. Our conclusions are as follows: firstly,
we note that the contributions to the anomalous dimension at this order
fall into two classes, proportional respectively to $F^{4}$ and 
$\pi^2F^{4}$, where
$F$ is the usual factor associated with loops in dimensional regularisation,
in 3 dimensions $F=\frac{1}{8\pi}$. 
The latter class has been called ``maximally transcendental''\cite{MSS, MSSa},
and we shall call the former ``rational''.  We then show that
the maximally-transcendental contributions
to the four-loop anomalous dimension in general fail to vanish 
upon imposing lowest-order superconformality and hence
require a coupling redefinition to restore superconformality. We shall consider
in some detail the case of multi-trace deformations where the non-vanishing
contributions where it is particularly clear that a redefinition will always
be required. 
On the other hand,
we shall show that (at least to leading order in 
$N$, $M$, and probably to all orders) the ``rational'' contributions to
the four-loop anomalous dimension do vanish, for a large class of theories,
once the lowest-order 
superconformality conditions are imposed.

Finally, we discuss the possibility that any non-vanishing redefinition 
required might be expressible in a simple general form, analogous to
experience in four dimensions; this turns out to 
require that the divergent contribution from a certain non-planar diagram, 
which we have been unable to compute explicitly, must take a certain value. 

The paper is organised as follows: in Section 2 we describe the general 
$\Ncal=2$ supersymmetric $U(N)\times U(M)$ 
Chern-Simons theory in three dimensions together
with various choices of superpotential and couplings which render it 
superconformal; in Section 3 
we describe our calculations and give our main results; and in Section 4 we
discuss the issue of coupling redefinitions from the standpoint of a more
general theory. Section 5
contains some brief conclusions, and we explain
our conventions and list various useful basic results and identities in an 
Appendix.

\section{$\Ncal=2$ Chern-Simons theory in three dimensions}
We consider an $\Ncal=2$ supersymmetric $U(N)\times U(M)$ Chern-Simons
theory with vector multiplets $V$, $\Vhat$ in the adjoint
representations of $U(N)$ and $U(M)$ respectively, and we write
\be
V^b{}_a=V^A(R_A)^b{}_a,\quad \Vhat^{\bhat}{}_{\ahat}=
\Vhat^A(\Rhat_A)^{\bhat}{}_{\ahat}, 
\ee 
where $R_A$, $A=1,\ldots N^2$ and $\Rhat_A$, $A=1,\ldots M^2$ are the 
generators for the fundamental representations of $U(N)$, $U(M)$
respectively.

The vector multiplets are coupled to chiral multiplets $(A^i)^a{}_{\ahat}$ 
and $(B_i)^{\ahat}{}_a$, $i=1,2$ in the $(N,\Mbar)$ and $(\Nbar,M)$ 
representations of the gauge group, respectively. 
The gauge matrices $R_A$ satisfy
\bea
[R_A,R_B]=&if_{ABC}R_C,\nn
\Tr(R_AR_B)=&\delta_{AB},
\label{Tdef}
\eea
with similar expressions for $\Rhat_A$ with structure constants $\fhat_{ABC}$.

The action for the theory can be written 
\be
S=S_{SUSY}+S_{GF}
\label{lag}
\ee
where $S_{SUSY}$ is the usual supersymmetric action\cite{ivanov}
\bea
S_{SUSY}&=&\int d^3x\int d^4\theta
\int_0^1dt\left\{K_1\Tr[\Dbar^{\alpha}(
e^{-tV}D_{\alpha}e^{tV})]+K_2\Tr[\Dbar^{\alpha}(
e^{-t\Vhat}D_{\alpha}e^{t\Vhat})]\right\}\nn
&+&\int d^3x\int d^4\theta\Tr\left(\Abar_ie^VA^ie^{-\Vhat}+
\Bbar^ie^{\Vhat}B_ie^{-V}\right)\nn
&+&\left(\int d^3x\int d^2\theta W(A^i,B_i)
+\hbox{h.c.}\right).
\label{ssusy}
\eea
Here the 
superpotential (quartic for renormalisability in three dimensions)
$W(A^i,B_i)$ is given by
\bea
W(A^i,B_i)&=&\Tr[h_1(A^1B_1)^2+h_2(A^2B_2)^2+h_3A^1B_1A^2B_2+h_4A^2B_1A^1B_2]\nn
&+&\frac12H_1[\Tr(A^1B_1)]^2
+H_{12}\Tr(A^1B_1)\Tr(A^2B_2)+\frac12H_2[\Tr(A^2B_2)]^2
\label{Wdef}
\eea
Gauge invariance requires $2\pi K_1$ and $2\pi K_2$ to be integers. 

A variety of interesting theories may be obtained by specialising
the superpotential in Eq.~(\ref{Wdef}) and the gauge group and associated
Chern-Simons levels in various ways. For 
\be
H_{1,2}=H_{12}=0, \quad 
h_1=h_2=\frac12\left(\frac{1}{K_1}+\frac{1}{K_2}\right), 
\quad h_3=\frac{1}{K_1},\quad h_4=\frac{1}{K_2},
\label{Nthree}
\ee
we obtain the $\Ncal=3$ superconformal theory described in Ref.~\cite{GT}.
Specialising to $K_1=-K_2=K$, so that $h_1=h_2=0$, $h_3=-h_4=h$,  we obtain
the $\Ncal=2$ ABJM/ABJ-like theories studied in Ref.~\cite{penatia}. 
In particular, for $h=\frac{1}{K}$ one obtains the $\Ncal=6$ 
superconformal ABJ theory and for $N=M$ the ABJM theory. 
Additional more general superconformal theories may be found by
solving the lowest order finiteness conditions (see later). Further 
superconformal theories may also be obtained by adding flavour 
matter\cite{penatib}.

We now consider the details of gauge fixing and quantisation for our 
Chern-Simons theory.
In each gauge sector we choose
a gauge-fixing term $S_{GF}$ in Eq.~(\ref{lag}) given by\cite{penatib} 
\be
S_{GF}=\frac{K}{2\alpha}\int d^3xd^2\theta\tr[ff]
-\frac{K}{2\alpha}\int d^3xd^2\thetabar\tr[\fbar\fbar]
\label{sgf}
\ee 
and we introduce into the functional integral a corresponding ghost term
\be  
\int{\cal D}f{\cal D}\fbar\Delta(V)\Delta^{-1}V
\ee
with 
\be
\Delta(V)=\int d\Lambda d \Lambdabar\delta(F(V,\Lambda,\Lambdabar)-f)
\delta(\Fbar(V,\Lambda,\Lambdabar)-\fbar),
\label{ghostdet}
\ee
with $\Fbar=D^2V$, $F=\Dbar^2V$.
With $\alpha=0$ this results in a gauge propagator for $V$ of the form
\be
\langle V^A(1)V^B(2)\rangle=-\frac1K_1\frac{1}{\pa^2}
\Dbar^{\alpha}D_{\alpha}\delta^4(\theta_1-\theta_2)\delta^{AB},
\ee
with a similar propagator for $\Vhat$.
The gauge vertices are obtained by expanding $S_{SUSY}+S_{GF}$ as given by
Eqs.~(\ref{ssusy}), (\ref{sgf}): 
\bea
S_{SUSY}+S_{GF}&\rightarrow&
-\frac{i}{6}K_1f^{ABC}\int d^3xd^4\theta\Dbar^{\alpha}V^A
D_{\alpha}V^BV^C\nn
&-&\frac{i}{6}K_2\fhat^{ABC}\int d^3xd^4\theta\Dbar^{\alpha}\Vhat^A
D_{\alpha}\Vhat^B\Vhat^C+\ldots\nn
\eea

The ghost action resulting from Eq.~(\ref{ghostdet}) has the same form
as in the four-dimensional $\Ncal=1$ case\cite{GGRS,GRS}
\be
S_{gh}=\int d^3xd^4\theta\tr\{\cbar'c-c'\cbar+
\frac12(c+\cbar')[V,c+\cbar)]+
\frac{1}{12}(c+\cbar')[V,[V,c-\cbar)]]+\ldots\}
\label{sgh}
\ee
leading to ghost propagators
\be
\langle \cbar'(1)c(2)\rangle=-\langle c'(1)\cbar(2)\rangle
=-\frac{1}{\pa^2}\delta^4(\theta_1-\theta_2),
\ee
(together with similar expressions involving $\Vhat$ and its own ghosts),
and cubic and higher-order vertices which may easily be read off from 
Eq.~(\ref{sgh}).
Finally the chiral propagator and chiral-gauge vertices are readily obtained
by expanding Eq.(\ref{ssusy}); the chiral propagators are given by:
\be
\langle \Abar_i^{\ahat}{}_a A^{jb}{}_{\bhat}\rangle
=-\frac{1}{\pa^2}\delta^4(\theta_1-\theta_2)
\delta_a{}^b\delta^{\ahat}{}_{\bhat}\delta^j{}_i,
\ee
with a similar expression for the $B$-propagator.

The regularisation of the theory is effected by replacing 
$V$, $\Vhat$, $A$, $B$, $h_i$, $H_i$ (and the various ghost fields)
by corresponding bare quantities 
$V_B$, $\Vhat_B$, $A_B$, $B_B$, $h_{Bi}$, $H_{Bi}$ (and similarly for the ghost 
fields) with the bare and renormalised fields related by  
\be
V_B=Z_V^{\frac12}V, \quad Z_A=Z_A^{\frac12}A,
\ee 
etc.
We use dimensional regularisation, working in $d$ dimensions with 
$d=3-2\epsilon$; so that divergences appear as poles in $\epsilon$.
The renormalisation constants $Z$ are power series in $\epsilon$ with
coefficients chosen to cancel poles in the two-point function at each loop
order. We can write (choosing $A^1$ as an example) 
\be
Z_{A^1}=\sum_LZ^{(L)}_{A^1}=\sum_{L\,\rm{even},m=1\ldots \frac{L}{2}}
\frac{Z_{A^1}^{(L,m)}}{\epsilon^m},
\label{Zbare}
\ee
where $L$ labels the loop order.
The corresponding anomalous dimensions such as $\gamma_{A^1}$ are defined by 
\be
\gamma_{A^1}=\mu\frac{d}{d\mu}\ln Z_{A^1},
\ee
where $\mu$ is the usual dimensional regularisation mass scale (introduced 
to preserve dimensions of couplings away from $d=3$ dimensions).
Using the fact that $Z_{A^1}$ on the left-hand side of Eq.~(\ref{Zbare})
is $\mu$-independent, while the renormalised couplings in $Z_{A^1}^{(L,m)}$ are 
$\mu$-dependent, implies that 
$\gamma_{A^1}$ is determined by the simple poles in $Z_{A^1}$ according
to
\be
\gamma_{A^1}^{(L)}=LZ_{A^1}^{(L,1)};
\label{gamdef}
\ee
and the 
higher order poles in $Z_{A^1}$ 
are determined by consistency conditions, the one
relevant for our purposes being 
\be
16Z_{A^1}^{(4,2)}
=2\left(\gamma_{A^1}^{(2)}\right)^2
-\sum_r\beta_{\lambda_r}^{(2)}.\frac{\pa}{\pa \lambda_r}\gamma_{A^1}^{(2)}
\label{doublep}
\ee
where $\{\lambda_r,r=1\ldots14\}=\{h_i, \barh_i, H_i, \Hbar_i, H_{12}, 
\Hbar_{12}\}$. The $\beta$-functions $\beta_{\lambda_r}$ in Eq.~(\ref{doublep})
are defined as usual by (picking $h_3$ for instance)
\be
\beta_{h_3}=\mu\frac{d}{d\mu} h_3
\ee
and measure the scale dependence of the renormalised couplings.
For a superconformal theory all the $\beta$-functions must therefore vanish. 
Since the $\beta$-functions for the Chern-Simons levels $K_{1,2}$ are 
expected to vanish for a 
generic Chern-Simons theory\cite{Kap} 
due to the topological nature of the theory (so that $K_{B1,2}=K_{1,2}$), 
superconformality will be determined purely by the vanishing of the 
$\beta$-functions for the superpotential couplings. For a general theory
the $\beta$-functions are
given in terms of the simple poles in the corresponding bare 
coupling, analogously to Eq.~(\ref{gamdef}). 
However for $\Ncal=2$ supersymmetric theories in three dimensions (as for
$\Ncal=1$ supersymmetric theories in four dimensions), the 
$\beta$-functions can be expressed 
according to the non-renormalisation theorem in terms of the 
anomalous dimensions of the fields associated with each coupling; for 
instance 
\be
\beta_{h_3}=(\gamma_{A^1}+\gamma_{B_1}+\gamma_{A^2}+\gamma_{B_2})h_3,
\label{nonren}
\ee 
with similar expressions
for the other superpotential couplings (the $\beta$-function for any coupling
is the same as that for its conjugate).
At lowest order (two loops) it was found that superconformality (i.e. 
the vanishing of $\beta_{\lambda_r}$) was equivalent to the vanishing of 
all the corresponding anomalous dimensions (for the fields involved in the 
$\lambda_r$ coupling)
in all the cases considered\cite{penatib, ASW}.

\section{Perturbative Calculations}
In this section we review the two-loop calculation and describe in 
detail our four-loop results.

The renormalisation constants of the chiral superfields $A^{1,2}$, $B_{1,2}$
are given 
at two loops by\cite{penatib,ASW}
\be
F^{-2}\gamma_{A^1}^{(2)}=2(\rho_{A^1}-\rho_k)
\label{Ztwo}
\ee
(with similar expressions for $A^2$, $B_{1,2}$) where $F=\frac{1}{8\pi}$
as defined before and
\bea
\rho_{A^1}=\rho_{B_1}&=&4|h_1|^2(MN+1)+(|h_3|^2+|h_4|^2)MN+(h_3\barh_4+h_4\barh_3)\nn
&+&
MN(|H_1|^2+|H_{12}|^2)+|H_1|^2,\nn
\rho_{A^2}=\rho_{B_2}&=&4|h_2|^2(MN+1)+(|h_3|^2+|h_4|^2)MN+(h_3\barh_4+h_4\barh_3)\nn
&+&MN(|H_2|^2+|H_{12}|^2)+|H_2|^2,\nn
\rho_k&=&(k_1^2+k_2^2)(2MN+1)+2(MN+2)k_1k_2,
\label{Gdef}
\eea  
with
\be
k_1=\frac{1}{K_1},\quad k_2=\frac{1}{K_2}.
\ee

\begin{figure}
\includegraphics{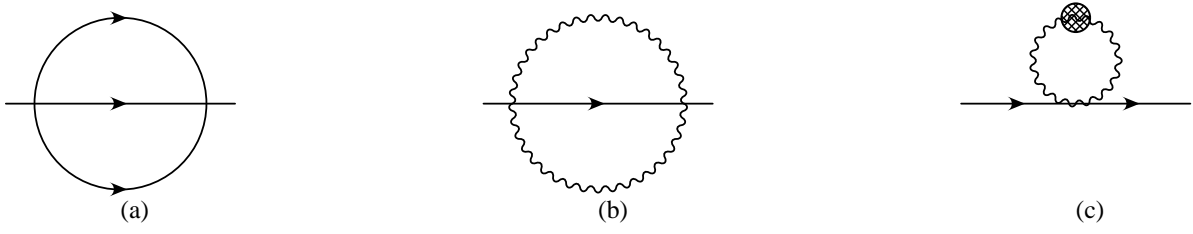}
\caption{Two-loop diagrams}
\label{twoloop}
\end{figure}

\begin{figure}
\includegraphics{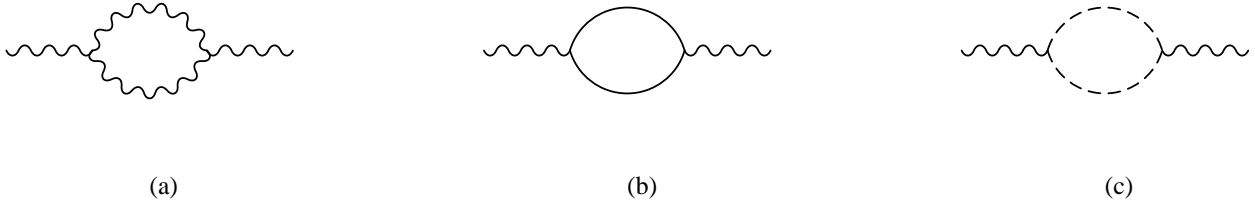}
\caption{One-loop insertion diagrams}
\label{oneloop}
\end{figure}

This result may readily be obtained by $\Ncal=2$ superfield 
methods\cite{agk,GN,ASW,penatib} from the two-loop two-point diagrams depicted
in Fig.~\ref{twoloop}; see
the Appendix for our $\Ncal=2$ superfield conventions. Here and later we do not
distinguish in the diagrams between the different chiral or gauge fields, 
so that each diagram in Fig.~\ref{twoloop} is a schematic representation of 
several distinct Feynman diagrams. 
$\rho_{A^1}$ etc correspond to Fig.~\ref{twoloop}(a) while 
it may easily be checked that
\be
\rho_k=\rho_b+\rho_c
\label{twog}
\ee
where the contributions $\rho_{b,c}$ corresponding to Fig.~\ref{twoloop}(b,c) 
are given by
\bea
\rho_b&=&\frac12(C_1+C_2)=\frac12[(N^2+1)k_1^2+(M^2+1)k_2^2+4MNk_1k_2],\nn
\rho_c&=&\frac12[X_1Nk_1^2+X_2Mk_2^2+X_{12}k_1k_2],
\label{twoga}
\eea
with
\bea
C_1&=&N^2k_1^2+M^2k_2^2+2MNk_1k_2,\nn
C_2&=&k_1^2+k_2^2+2MNk_1k_2,\nn
X_1&=&4M-\frac{N^2-1}{N},\nn
X_2&=&4N-\frac{M^2-1}{M},\nn
X_{12}&=&8.
\label{twogb}
\eea
$C_{1,2}$ correspond to the two different symmetrisations of the gauge 
lines in Fig.~\ref{twoloop}(b), while the $X_{1,2}$, $X_{12}$ correspond 
to the contributions from the ``blob'' in Fig.~\ref{twoloop}(c) which
represents the three one-loop diagrams depicted in Fig.~\ref{oneloop}
(the dashed line representing a ghost propagator). 

(We note here that the two-loop
results for general Chern-Simons theories obtained in 
Ref.~\cite{akk} are not directly comparable since they were computed
in the $N=1$ framework.) 

As mentioned in the Introduction, we shall consider two classes of model in 
some detail; the first without, and the second with, 
multitrace deformations. We shall call these Class I, Class II theories 
respectively. Class I corresponds to taking   
$H_{1,2}=H_{12}=0$ in Eq.~(\ref{Wdef}); and in fact we start with the 
even simpler example of   
\be
H_{1,2}=H_{12}=0, \quad h_1=h_2=0, \quad h_3+h_4=0,
\label{caseone}
\ee
with $h_3=-h_4=h$ real; we shall call this Class Ia. 
This is a class of theories considered in 
Ref.~\cite{GT}, which reduces to the ABJ model on setting 
$K_1=-K_2$ (or $k_1=-k_2$) and to the ABJM model on further setting $M=N$.

\begin{figure}
\includegraphics{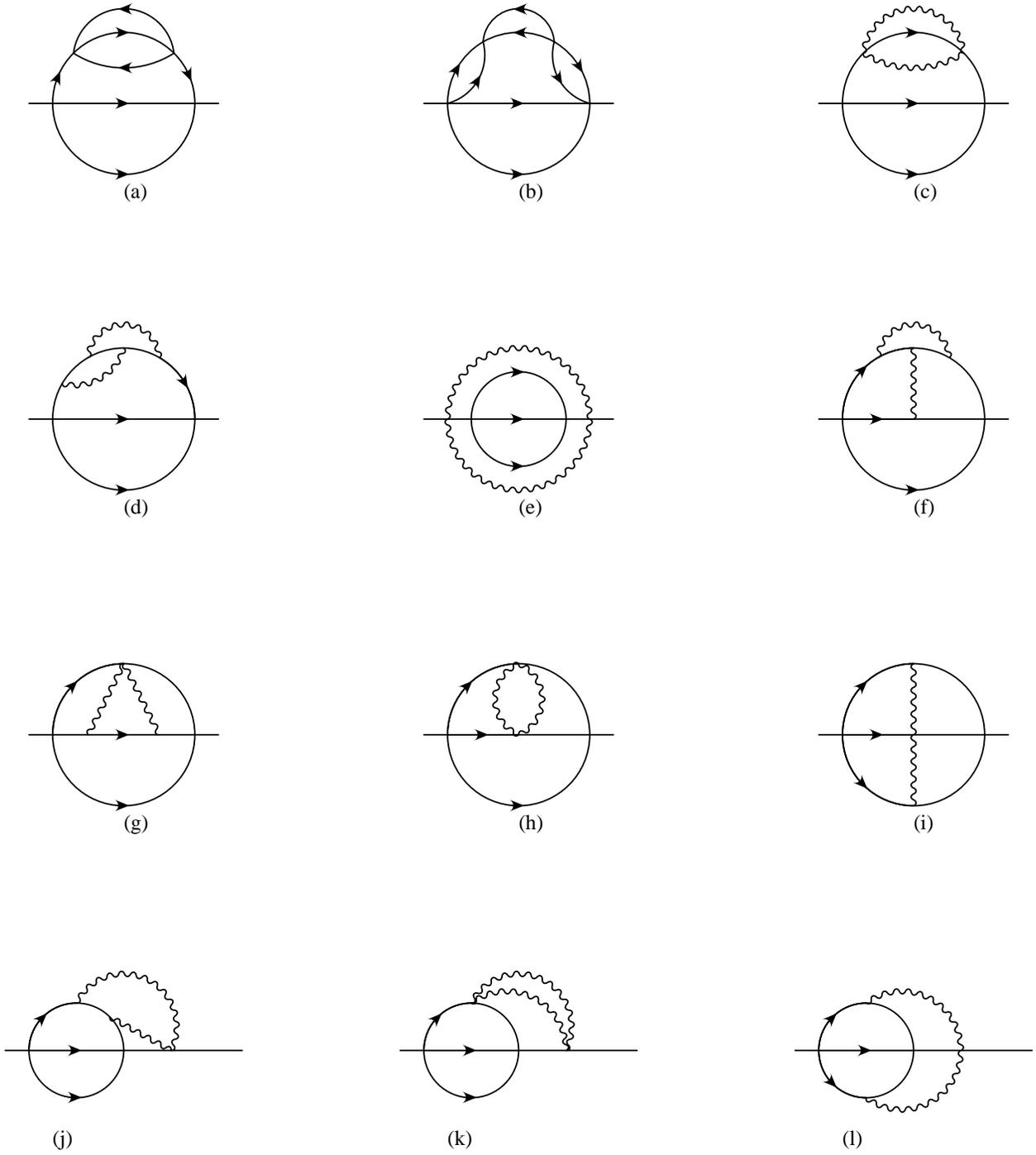}
\caption{Four-loop diagrams}
\label{diags}
\end{figure}

\begin{figure}
\includegraphics{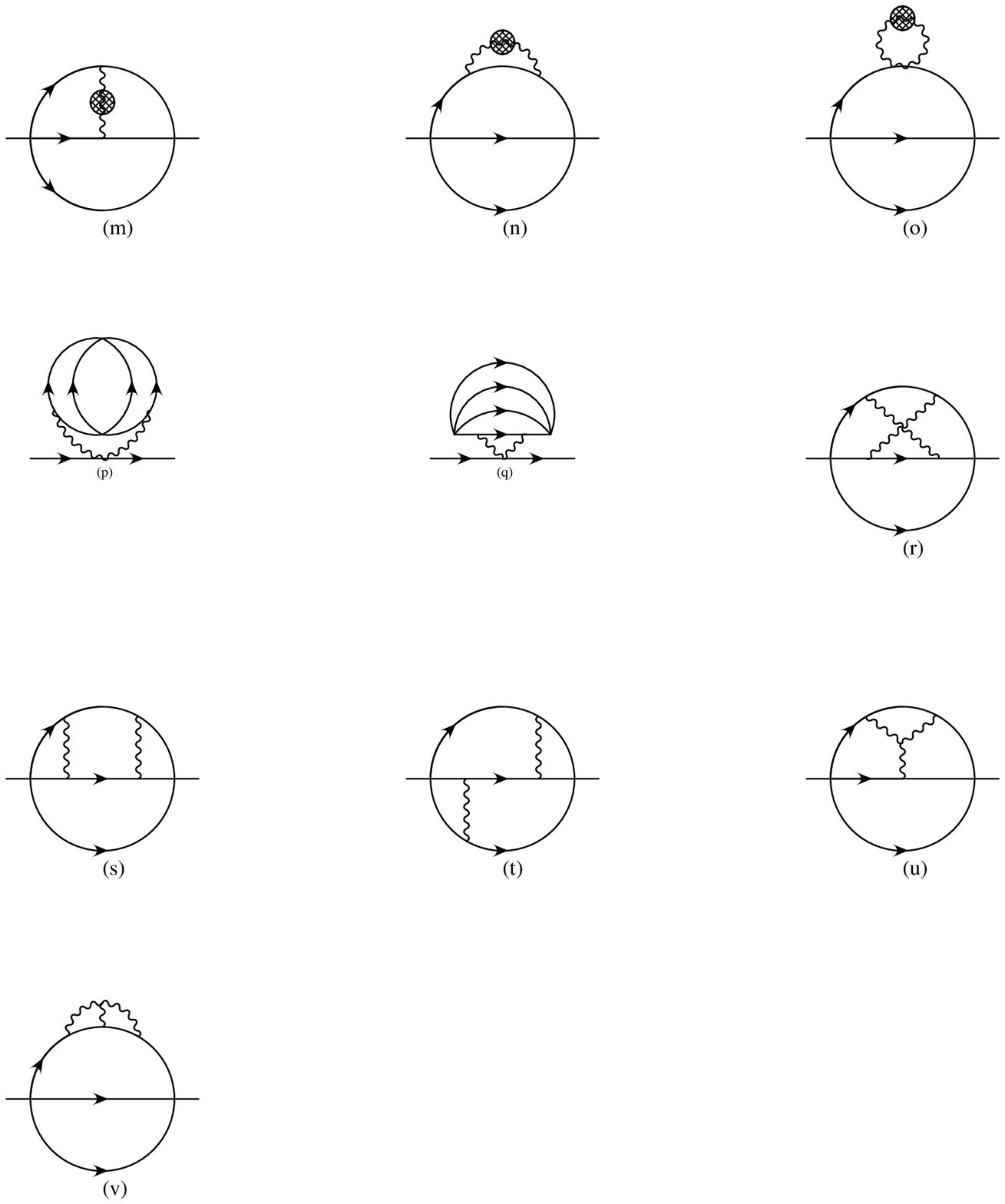}
\caption{Four-loop diagrams (continued)}
\label{diagsa}
\end{figure}

The four-loop diagrams with Yukawa couplings contributing
to the anomalous dimensions are depicted in Figs.~\ref{diags}, \ref{diagsa}.
As we explained before, we have not considered the much larger set of diagrams
with no Yukawa couplings, but we shall still be able to draw some conclusions.
The contributions to $F^{-4}Z^{(4)}_{A^1}$ from these diagrams are given by
\bea
G_a&=&3\rho_h^2I_4,\nn
G_b&=&2(MN^3+NM^3-4M^2-4N^2+10MN-4)h^4I_{4bbb},\nn
G_c&=&-3\rho_h\rho_bI_4,\nn
G_d&=&3\rho_hC_2I_{4bbb},\nn
G_e&=&-\rho_h\rho_bI_4,\nn
G_f&=&2\rho_hC_2I_5,\nn
G_g&=&-T_2I_5,\nn
G_h&=&-T_2I_{4bbb},\nn
G_i&=&-T_1I_5,\nn
G_j&=&4T_2\left(I_4-\frac12I_{4bbb}\right),\nn
G_k&=&-2T_2I_4,\nn
G_l&=&-2T_1(-2I_4+2I_{4bbb}+I_{5}),\nn
G_m &=&-2\rho_h\rho_c\left(I_4-\frac12I_{4bbb}\right),\nn
G_n&=&3\rho_h\rho_cI_{22},\nn
G_o&=&-3\rho_h\rho_cI_{22},\nn
G_p&=&2\rho_hT_3(I_4-J_4-I_{42bbc}),\nn
G_q&=&-2\rho_hT_3(I_4-J_4),\nn
G_r&=&\frac{1}{\epsilon}T_4(a+b\pi^2)
\label{fourres}
\eea
where
\be
\rho_h=2(MN-1)h^2
\label{Ghdef}
\ee
is the common value of $\rho_{A^{1,2}}$, $\rho_{B_{1,2}}$ upon imposing 
Eq.~(\ref{caseone}) and
\bea
T_1&=&h^2[(N^2-2MN+1)k_1^2+(M^2-2MN+1)k_2^2\nn
&+&2(M^2N^2+2MN-M^2-N^2-1)k_1k_2],\nn
T_2&=&h^2[(N^3M+5MN-3N^2-3)k_1^2+(M^3N+5MN-3M^2-3)k_2^2\nn
&+&4(M^2+N^2-3MN+1)k_1k_2],\nn
T_3&=&4[(k_1^2+k_2^2)MN+2k_1k_2],\nn
T_4&=&h^2[(3MN-N^2-2)k_1^2+(3MN-M^2-2)k_2^2\nn
&+&2(N^2+M^2-3MN+1)k_1k_2].
\label{Tedef} 
\eea
These quantities are not all independent and in fact (as a consequence of
gauge invariance) satisfy the identities
\bea
2T_1+T_2&=&\rho_h\rho_b,\nn
T_1+T_4&=&\frac12\rho_hC_2,
\label{Trel}
\eea
where $\rho_b$, $C_2$ and $\rho_h$ are defined in Eqs.~(\ref{twoga}),
(\ref{twogb}) and (\ref{Ghdef}).
The results are expressed in terms of a basis of momentum integrals 
defined and computed in Ref.~\cite{MSS}. The divergent contributions from
these momentum integrals are listed in the Appendix. We have not been able to compute the 
momentum integral corresponding to Fig.~\ref{diagsa}(r), and therefore $a$, $b$ in 
$G_r$ in Eq.~(\ref{fourres}) are unknown. This momentum integral is depicted in 
Fig.~\ref{allmom}(g) in the Appendix.
The contributions from
Fig.~\ref{diagsa}(s)-(v) are all finite or zero and therefore not listed 
explicitly.

The full result obtained by summing the individual contributions in 
Eq.~(\ref{fourres}):
\bea
F^{-4}Z_{A^1}^{(4)}&=&G^{(4)},\nn
G^{(4)}&=&G_a+\ldots+G_r+R,
\label{Zdef}
\eea
where $R$ represents the (currently unknown) contribution from graphs with no
Yukawa couplings, 
may be divided into transcendental and rational contributions (according 
to whether the contribution contains  a factor of $\pi^2$ or not, 
coming from the right-hand sides of Eqs.~(\ref{momint})
) as
\be
G^{(4)}=G_{\hbox{rat}}+G_{\hbox{trans}}\pi^2
\ee
The transcendental contribution is given (using Eq.~(\ref{Trel})) by
\bea
G_{\hbox{trans}}
&=&\frac{1}{\epsilon}\Bigl\{h^4(MN^3+NM^3-4M^2-4N^2+10MN-4)\nn
&+&\frac16[(5+3b)\rho_hC_2+3\rho_h\rho_k-10\rho_h\rho_b+2(4-3b)T_1]\Bigr\}
+R_{\hbox{trans}},
\label{Gfulla}
\eea
with obvious definitions for $R_{\hbox{trans}}$, $R_{\hbox{rat}}$.
We shall postpone comment on this until later, and focus on the rational 
contribution, which is given (again using Eq.~(\ref{Trel})) by
\be
G_{\hbox{rat}}=3\rho_h^2I_4-2\rho_h\rho_kI_4-2\rho_hT_3I_{42bbc}
+\frac{a}{\epsilon}T_4+R_{\hbox{rat}}
\label{Grat} 
\ee
where $\rho_k$ is given by Eq.~(\ref{Gdef}). We have used here the fact that
$I_5$ as defined in Eq.~(\ref{momint}) gives only a transcendental simple pole.
Since $T_4$ is $O(N^2)$, the $a$ 
term from the non-planar graph $G_r$ certainly gives no contribution at
leading order $O(N^4)$; and based on experience with non-planar graphs, we
believe it is likely that $G_r$ gives a purely transcendental divergent 
contribution and hence $a=0$. 
Upon imposing the two-loop superconformality condition $\gamma^{(2)}=0$
in the form 
\be
\rho_h=\rho_k,
\label{conftwo}
\ee 
(using Eqs.~(\ref{Ztwo}), (\ref{Gdef}), (\ref{Ghdef})),  we find
\be
G_{\hbox{c,rat}}=\rho_k^2I_4-2\rho_kT_3I_{42bbc}+\frac{a}{\epsilon}T_4
+R_{\hbox{rat}},
\label{rat}
\ee
where $G_c$ denotes the value of $G$ upon imposing leading-order
superconformality. 
The value of $h=h_c(k_1,k_2)$ implicit in the 3rd term in Eq.~(\ref{rat}) 
(according to Eq.~(\ref{Tedef})) will be 
determined by solving Eq.~(\ref{conftwo})
and clearly depends on the particular form of the superpotential. 
However the remaining terms in Eq.~(\ref{rat}) are independent of $h$ and
thus (since we see from Eq.~(\ref{Tedef}) that 
the 3rd term is subleading in $N$, $M$) the form of 
$G_{\hbox{rat}}$ is independent of the form of the
superpotential to leading order.  
In fact, it is straightforward to see that this result is
more general and applies to any Class I theory, as we shall proceed to show.  
Firstly, the $I_4$ terms in Eq.~(\ref{fourres}) 
supply the double pole contributions of the form $h^4$ and $h^2k^2$; and 
this will remain the case for any Class I theory.
The $h^4$ terms are given according to Eqs.~(\ref{doublep}), (\ref{Gdef}) 
and (\ref{Ztwo}) by
\bea
&&\frac12\rho_{A^1}^2-4(\rho_{A^1}+\rho_{B_1})h_1^2(MN+1)\nn
&-&\frac12(\rho_{A^1}+\rho_{B_1}+\rho_{A^2}+\rho_{B_2})
[(|h_3|^2+|h_4|^2)MN+h_3\barh_4+h_4\barh_3]
 \eea
which reduces to $-\frac32\rho_k^2$ upon imposing 
the two-loop superconformal invariance condition, now from
Eqs.~(\ref{Ztwo})
\be
\rho_{A^{1,2}}=\rho_{B_{1,2}}=\rho_k.
\label{genconf}
\ee
This reproduces exactly the contribution of the first term in Eq.~(\ref{Grat})
to Eq.~(\ref{rat}).
The $h^2k^2$ terms are given according to Eq.~(\ref{doublep}) by
$\rho_k\rho_{A^1}$ which of course reduces to $\rho_k^2$ upon 
imposing $\rho_{A^1}=\rho_k$. This reproduces exactly the contribution of the 
second term in Eq.~(\ref{Grat})
to Eq.~(\ref{rat}). Furthermore, 
in the general case, the coefficient in $G_p$ in Eq.~(\ref{fourres})
becomes 
\be
-2(\rho_{A^1}+\rho_{A^2}+\rho_{B_1}+\rho_{B_2})[(k_1^2+k_2^2)MN+2k_1k_2]
\ee
which reduces to
\be
8\rho_k[(k_1^2+k_2^2)MN+2k_1k_2]=2\rho_kT_3 
\ee
upon imposing Eq.~(\ref{genconf}); now reproducing the
contribution of the third term in Eq.~(\ref{Grat})
to Eq.~(\ref{rat}). 
Finally, the
contribution from the non-planar graph $G_r$ is subleading in $N$, $M$ 
for any theory with superpotential of the form Eq.~(\ref{Wdef}) with
$H_{1,2}=H_{12}=0$; in fact the only reason we have had to exclude
multi-trace deformations from the definition of Class I is that otherwise 
this is no longer true. Therefore the form of 
$G_{\hbox{rat}}$ in Eq.~(\ref{rat}) is in general independent of the form of 
the potential at leading order in $M$, $N$ upon imposing the conformal 
invariance condition, as long as multi-trace deformations are excluded. Since 
we believe it likely that $a=0$, this result may well also hold at lower orders 
and in the presence of multi-trace deformations.
  
The results from the remaining diagrams with no Yukawa couplings are of course
also independent of the form of the potential, and 
the rational contribution from these graphs must take the form
\be
R_{\hbox{rat}}=-\rho_k^2I_4+\frac{1}{\epsilon}\delta(k_1,k_2)
\label{allk}
\ee
in order that the total double pole contribution to $G_{c,\hbox{rat}}$ cancels, 
as it must due to the lower order superconformal invariance. 
We therefore obtain from Eqs.~(\ref{Zdef}), (\ref{rat}),
(\ref{allk}) 
\be
F^{-4}Z_{\hbox{rat}A^1}^{(4)}=-2\rho_kT_3I_{42bbc}+\frac{a}{\epsilon}T_4
+\frac{1}{\epsilon}\delta(k_1,k_2). 
\ee
We are left (using Eqs.~(\ref{gamdef}), (\ref{momint})) with an expression
for the residual four-loop rational contribution to the anomalous dimension 
after imposing two-loop superconformal invariance, valid for any Class I 
theory (and any field, so we therefore suppress the field label):
\be
F^{-4}\gamma_{\hbox{c,rat}}^{(4)}=-16\rho_kT_3+4a{\tilde T}_4+4\delta(k_1,k_2).
\label{indep}
\ee
Here $\tilde T_4$ represents the generalisation to a general Class I 
theory of the expression in Eq.~(\ref{Tedef}), which will now
depend on $h_{1-4}$. 
Now we know that $\gamma_{\hbox{rat}}^{(4)}$ 
must vanish when $h_{1-4}$ take the values given in Eq.~(\ref{Nthree})
corresponding to $\Ncal=3$ supersymmetry, since this theory is superconformal
without any renormalisation\cite{GY}. However in Eq.~(\ref{indep}),
$h_{1-4}$ only appear in the $a$ term, and in particular in the 
subleading part of ${\tilde T}_4$.
If $a=0$, we can immediately deduce
that $\delta(k_1,k_2)=4\rho_k T_3$ and hence (since this conclusion is
independent of the values of $h_{1-4}$) that $\gamma_{\hbox{c,rat}}^{(4)}=0$
for any Class I theory. However, if $a\ne0$, the most we can say 
is that the leading term in  
$\gamma_{\hbox{c,rat}}^{(4)}$ vanishes for any Class I theory; but a 
``rational''
coupling redefinition may in general be required to restore superconformal 
invariance for the subleading terms--see later for the general form of this
redefinition.

We believe that 
our result will also extend to the superconformal theories with flavour matter
discussed in Ref.~\cite{penatib}; and, if $a=0$ in $G_r$ in Eq.~(\ref{fourres}),
to theories with multi-trace deformations as well. We would be able to
apply the same arguments in the case with flavour as in the situation just 
discussed, since the wide class of
theories with flavour discussed in Ref.~\cite{penatib} contains an 
$\Ncal=3$ theory with flavour as a special case for particular choices of 
coupling; and once again this $\Ncal=3$ theory is exactly 
superconformal\cite{GJ}.

We shall not consider further here the transcendental contribution 
for the Class I models, since we 
can draw a more striking conclusion in the case of the Class II
models; suffice it to 
say that the expression given in Eq.~(\ref{Gfulla}) for the Class 1a
models clearly gives a model-dependent result upon 
imposing two-loop superconformality, Eq.~(\ref{conftwo}) (in that the $h^4$ 
terms, and the $h^2$ terms in $T_1$, are a consequence of the choice of
superpotential). Therefore
although $G_{\hbox{trans}}$ must vanish for $h=k=k_1=-k_2$ 
(corresponding to $\Ncal=3$ supersymmetry) it cannot vanish for general
$h$ satisfying Eq.~(\ref{conftwo}) and hence a ``transcendental'' coupling 
redefinition will be required to restore superconformality.

Before leaving the Class I models, we point out that by using $\Ncal=3$
superconformality we have determined the ``rational" remainder term 
$R_{\hbox{rat}}$ in Eq.~(\ref{Grat}), up to the unknown value of $a$; and
we could also obtain in a similar way the ``transcendental" remainder term 
$R_{\hbox{trans}}$ in Eq.~(\ref{Gfulla}), up to the unknown value of $b$,
and for $k_2=-k_1$. We have tried to derive our conclusions with the minimum
effort; but one could straightforwardly extend our calculation to the
case of general $h_{1-4}$, whereupon imposing $\Ncal=3$ supersymmetry
would enable one to derive $R_{\hbox{trans}}$ for general $k_{1,2}$ 
($b$ of course remaining as an unknown).

We now turn to the Class II models, containing 
multi-trace deformations. We consider the simplest example of such a model, 
taking in Eq.~(\ref{Wdef})
\be
M=N,\quad k_1=-k_2=k,\quad h_3=-h_4=h,\quad
H_{12}=H_1=H_2=H.
\ee
In this case the two-loop result in Eq.~(\ref{Ztwo}) reduces to
\be
F^{-2}\gamma^{(2)}_{A^1}=2(\rho_H-\rho_k)
\label{ZHtwo}
\ee
where
\be
\rho_H=2h^2(N^2-1)+H^2(2N^2+1)
\label{GHdef}
\ee
with $\rho_k$ given according to Eq.~(\ref{twoga}) but with now in 
Eq.(\ref{twogb}) 
\be
C_1=0,\quad C_2=-2(N^2-1)k^2,
\label{Cpdef}
\ee
so that 
\bea
\rho_b=-(N^2-1)k^2, &\quad& \rho_c=3(N^2-1)k^2,\nn 
\rho_k&=&2(N^2-1)k^2.
\label{GkHdef}
\eea
The results for the diagrams in Figs~\ref{diags}, \ref{diagsa} are now given by
\bea
G'_{a}&=&3\rho_H^2I_4,\nn
G'_{b}&=&[4(N^2-1)(N^2+2)h^4+36(N^2-1)h^2H^2\nn
&+&(2N^2+1)(4N^4+6N^2+5)H^4]
I_{4bbb},\nn
G'_c&=&-3\rho_b\rho_HI_4,\nn
G'_d&=&3C_2\rho_HI_{4bbb},\nn
G'_e&=&-\rho_b\rho_HI_4,\nn 
G'_{f}&=&2C_2\rho_HI_5,\nn
G'_{g}&=&-T'_{2}I_5,\nn
G'_{h}&=&-T'_{2}I_{4bbb},\nn
G'_{i}&=&-T'_{1}I_5,\nn
G'_{j}&=&4T'_{2}\left(I_4-\frac12I_{4bbb}\right),\nn
G'_{k}&=&-2T'_{2}I_4,\nn
G'_{l}&=&-2T'_{1}(-2I_4+2I_{4bbb}+I_{5}),\nn
G'_{m} &=&-2\rho_H\rho_c\left(I_4-\frac12I_{4bbb}\right),\nn
G'_n&=&3\rho_c\rho_HI_{22},\nn
G'_o&=&-3\rho_c\rho_HI_{22},\nn
G'_p&=&2\rho_HT_3'(I_4-J_4-I_{42bbc}),\nn
G'_q&=&-2\rho_HT_3'(I_4-J_4),\nn
G'_r&=&\frac{1}{\epsilon}T'_4(a+b\pi^2)
\label{graphsa}
\eea
where
\bea
T'_1&=&-(N^2-1)k^2[2(N^2+2)h^2+3H^2],\nn
T'_2&=&(N^2-1)k^2[2(N^2+5)h^2-(2N^2-5)H^2],\nn
T'_3&=&8(N^2-1)k^2,\nn
T'_4&=&2(N^2-1)[3h^2-(N^2-1)H^2]k^2.
\eea
The quantities $T'_{1,2,4}$ satisfy identities similar to Eq.~(\ref{Trel}), 
namely
\bea
2T'_1+T'_2&=&\rho_H\rho_b,\nn
T'_1+T'_4&=&\frac12\rho_HC_2,
\label{Tprel}
\eea
but where $\rho_H$, $C_2$ and $\rho_b$
are now as defined in Eqs.~(\ref{GHdef}), (\ref{Cpdef}) and (\ref{GkHdef}).
The case $M=N$ and 
$k_1=-k_2$ can be expressed in terms of the 3-algebra formalism\cite{ASW}; 
this lends itself to automation and the results in Eq.~(\ref{graphsa}) were 
obtained using 
{\sc Form}\cite{form}.

For this class of models we shall start by discussing the
transcendental contributions to the anomalous dimension, since the results
are more striking than for the rational case. The
transcendental contribution is given by summing the 
contributions involving $I_{4bbb}$ and $I_5$ in Eq.~(\ref{graphsa}) 
together with $G'_r$ and 
using Eqs.~(\ref{Tprel}), (\ref{momint}) 
(and including the contribution $R_{\hbox{trans}}$
from graphs with no Yukawa couplings):
\bea
G_{\hbox{trans}}&=&
\frac{1}{2\epsilon}\Bigl\{4(N^2-1)(N^2+2)h^4+36(N^2-1)h^2H^2
\nn
&+&(2N^2+1)(4N^4+6N^2+5)H^4\nn
&+&\frac13[(5+3b)\rho_HC_2+3\rho_H\rho_k-10\rho_H\rho_b+2(4-3b)T'_1]\Bigr\}
+R_{\hbox{trans}}.
\label{transH}
\eea
To lowest order the vanishing of the anomalous dimensions now requires
(using Eqs.~(\ref{ZHtwo}), (\ref{GHdef}), (\ref{GkHdef}))
that the couplings $h$ and $H$ must be chosen to satisfy
\be
F^{-2}\gamma^{(2)}=2[2(N^2-1)h^2+(2N^2+1)H^2-2(N^2-1)k^2]=0.
\label{confH}
\ee
(We suppress the field label $A^{1,2}$, $B_{1,2}$ on $\gamma$ here and 
later, since for this class
of models every field has the same anomalous dimension.)
In order for $G_{\hbox{trans}}$ to adopt a universal form upon imposing
two-loop superconformal invariance as in Eq.~(\ref{confH}), we would require
Eq.~(\ref{transH}) to adopt the form
\be
G_{\hbox{trans}}=f(2(N^2-1)h^2+(2N^2+1)H^2).
\ee
This is clearly not the case. We shall therefore consider the two cases 
$H=0$ and $H\ne0$ separately, and find that they are very different.
If $H=0$ (so that we are considering the ABJM model) then Eq.~(\ref{transH}) 
reduces to
\be
G_{\hbox{trans}}=\frac{1}{\epsilon}(N^2-1)
\left[2(N^2+2)h^4-\frac23[N^2+11-9b]h^2k^2\right]+R_{\hbox{trans}}.
\ee
For $H=0$ the lowest-order superconformal invariance condition $\gamma^{(2)}=0$ 
simply implies (see Eq.~(\ref{confH})) that  
\be
h=h_c=k.
\label{confh}
\ee
If we assume
\be
R_{\hbox{trans}}=\frac{1}{\epsilon}(N^2-1)R_kk^4
\label{gaugecont}
\ee
then using Eq.~(\ref{gamdef})
the total transcendental contribution to the anomalous dimension at this
order is
\be
F^{-4}\gamma_{\hbox{trans}}^{(4)}=4(N^2-1)\sigma
\label{gamfourt}
\ee
where
\be
\sigma=2(N^2+2)h_c^4-\frac23(N^2+11-9b)h_c^2k^2+R_kk^4.
\label{sigdef}
\ee
This is easily derived by using Eqs.~(\ref{GHdef}), (\ref{Cpdef}), 
(\ref{GkHdef}) in Eq.~(\ref{transH}), in conjunction with Eq.~(\ref{ZHtwo}).
(The factor $N^2-1$ 
in Eq.~(\ref{gaugecont}) may be inferred from the fact that for $K_1=-K_2$,
all the contributions vanish identically in the abelian case due to the 
``quiver'' structure.) We have refrained from making the replacement
$h_c=k$ in Eq.(\ref{sigdef}) for later reference.  Now since we know that the 
ABJM model is exactly superconformal when $h_c=k$, we can deduce that
$\sigma$ must vanish for this value of $h_c$ and so
\be
R_k=-\frac23(2N^2-5+9b).
\label{Rres}
\ee
Returning to the case 
of $H\ne0$, it is clear that the exact superconformal invariance will 
no longer persist; if we substitute the superconformal invariance condition 
in the form $\rho_H=\rho_k$ into Eq.~(\ref{transH}) and use Eq.~(\ref{Rres}),
we obtain a non-vanishing result for $G_{c,\hbox{trans}}$
and hence for the transcendental contribution to $\gamma_c^{(4)}$.  
We write this as
\bea
\gamma^{(4)}_{c,\hbox{trans}}
&=&\alpha_1h_c^4+\alpha_2h_c^2H_c^2+\alpha_3H_c^4
+(\alpha_4h_c^2+\alpha_5H_c^2)k^2+\alpha_6(N^2-1)k^4,
\nn
\alpha_1&=&8(N^2-1)(N^2+2),\nn
\alpha_2&=&72(N^2-1),\nn
\alpha_3&=&2(2N^2+1)(4N^4+6N^2+5),\nn
\alpha_4&=&\frac83(3b-4)(N^2-1)(N^2+2),\nn
\alpha_5&=&4(3b-4)(N^2-1),\nn
\alpha_6&=&\frac83(1-3b)(N^2+2),
\label{classtwo}
\eea
where $h_c$ and $H_c$ together are solutions of Eq.~(\ref{confH}). 
Since $\alpha_3$ is $O(N^6)$ while all other
terms at this loop order are at most $O(N^4)$, it is particularly clear that 
we need to make a redefinition in Eq.~(\ref{ZHtwo})
to restore superconformal invariance at this order for this theory. A 
change $\delta h_c$,
$\delta H_c$ produces a change in $\gamma_c^{(2)}$ given according to 
Eq.~(\ref{ZHtwo}) by 
\be
F^{-2}\delta \gamma_c^{(2)}=4[2(N^2-1)h_c\delta h_c+(2N^2+1)H_c\delta H_c]
\ee
and we can therefore cancel the non-zero terms in 
$\gamma_{c,\hbox{trans}}{(4)}$ by taking    
\bea
F^{-2}\delta_{\hbox{trans}} h_c&=&-\frac{1}{8(N^2-1)}(\alpha_1h_c^2+\alpha_4k^2)h_c-
\kappa\frac{1}{8(N^2-1)}\alpha_2h_cH_c^2-\frac{1}{8}Ck^2h_c,\nn
F^{-2}\delta_{\hbox{trans}} H_c&=&-\frac{1}{4(2N^2+1)}(\alpha_3H_c^2+\alpha_5k^2)H_c-
(1-\kappa)\frac{1}{4(2N^2+1)}\alpha_2h_c^2H_c\nn
&&-\frac{1}{8}Ck^2H_c,
\label{redefH}
\eea
where $h_c$ and $H_c$ together solve Eq.~(\ref{confH})), and $\kappa$
is arbitrary. For the $\alpha_6$ terms we have again applied Eq.~(\ref{confH})).
Note that in Eq.~(\ref{redefH}) we are still suppressing
the ``transcendental'' factor of $\pi^2$.
We therefore conclude that 
a ``transcendental'' redefinition is  
inevitably required as soon the ABJM model is supplemented  by
multitrace deformations. We note
that Eqs.~(\ref{classtwo}), (\ref{redefH}) (with $H_c=0$)
could easily be adapted to
construct the redefinitions required in the Class I case.  
Finally turning to the rational contribution to the anomalous dimension for 
these models, the discussion would largely follow that for the previous class
of models. However as mentioned there, for $H_{1,2},H_{12}\ne0$ (and unless
$a=0$)
we would find a model-dependent contribution from $G_r'$ upon imposing
two-loop superconformal invariance, and this would require a model-dependent
redefinition of $H_{1,2},H_{12}\ne0$ akin to Eq.~(\ref{redefH}).

\bigskip
\section{General coupling redefinitions}
In this section we set the discussion of coupling redefinitions in a more
general context. We have postponed this discussion until now,
since it was easier for the purposes of exposition to define our models of 
interest explicitly {\it ab initio} than to start with a general theory and then
specialise. In this section we shall maintain the discussion at a general level
throughout without including too much detail.
  
In a general supersymmetric theory in three dimensions
with chiral fields $\Phi_i$, a general renormalisable 
superpotential would take the form
\be
W(\Phi)=Y^{ijkl}\Phi_i\Phi_j\Phi_k\Phi_l.
\ee
To lowest order the change in the $\beta$-function 
\be
\beta_Y^{ijkl}=\mu\frac{d}{d\mu}Y^{ijkl}
\ee
resulting from a change $\delta Y^{ijkl}$
is given by
\be
\delta\beta_Y^{ijkl}=\left(\beta_Y.\frac{\pa}{\pa Y}\right)\delta Y^{ijkl}
-\left(\delta Y.\frac{\pa}{\pa Y}\right)\beta_Y^{ijkl},
\ee
where
\be
\delta Y.\frac{\pa}{\pa Y}\equiv\delta Y^{mnpq}\frac{\pa}{\pa Y^{mnpq}}.
\ee
In the case of superconformal invariance at lowest order (we shall assume as 
before that this is equivalent to $\gamma=0$), this reduces to
\be
\delta\beta_Y^{ijkl}=\delta\gamma^{(i}{}_mY^{jkl)m}
\ee
where
\be
\delta\gamma^i{}_j=\delta Y.\frac{\pa}{\pa Y} \gamma^i{}_j.
\label{delgam}
\ee
The two-loop anomalous dimension is of the form 
\be
\gamma^{(2)i}{}_j=\frac13Y^{iklm}Y_{jklm}-C^i{}_j
\label{gengam}
\ee
with the convention that $Y_{ijkl}=(Y^{ijkl})^*$ and where $C^i{}_j$ is a 
function only of the Chern-Simon level(s). 
After imposing lowest-order superconformal invariance,
$O(k^4)$ terms in the four-loop anomalous dimension 
may be removed by a redefinition   
\be
\delta Y^{ijkl}=\frac12\lambda Y_c^{ijkl}
\label{geny}
\ee
where $Y_c^{ijkl}$ is a solution of the lowest order superconformal invariance
condition $\gamma^{(2)i}{}_j=0$ so that from Eq.~(\ref{gengam}) 
\be
\frac13Y_c^{iklm}Y_{cjklm}=C^i{}_j
\ee
and therefore using Eqs.~(\ref{delgam}), (\ref{geny})
\be
\delta\gamma^{(2)i}{}_j=\lambda C^i{}_j.
\ee
The redefinition 
in Eq.~(\ref{redefH}) which removes the $\alpha_6$ term in 
Eq.~(\ref{classtwo}) is clearly of this type. 

In terms of the general superpotential, the contribution to the
anomalous dimension from Fig.~\ref{diags}(b) (corresponding to $G_b$, $G_b'$
in Eqs.~(\ref{fourres}), (\ref{graphsa}) respectively)
is 
\be
\frac14\pi^2Y^{iklr}Y_{klmn}Y^{mnpq}Y_{pqrj}
\ee
and this may be removed completely by a redefinition 
\be
\delta Y^{ijkl}=-\frac18\pi^2Y^{mn(ij}Y_{mnpq}Y^{kl)pq}.
\label{genyy}
\ee
This reproduces (for a particular value of $\kappa$) 
the effect of the terms cubic in $h_c$, $H_c$ 
in Eq. (\ref{redefH}) in removing the quartic ($\alpha_1$, $\alpha_2$) terms in
Eq.~(\ref{classtwo}). We also note that if $b=\frac43$ 
(thus removing the $T_1'$
term in Eq.~(\ref{transH}) and setting $\alpha_4=\alpha_5=0$) then no further 
redefinition is required; whereas if $b\ne\frac43$ there is no such simple 
general form for the $\alpha_4$, $\alpha_5$ terms in Eq.~(\ref{redefH}).

We saw earlier that redefinitions such as Eq.~(\ref{genyy}) are not universal
in the sense that their form depends on the nature of the superpotential. 
Nevertheless it would be satisfying if all the necessary redefinitions  
could be expressed in a general form such as Eqs.~(\ref{geny}), (\ref{genyy}).
This would point to the existence of a ``superconformal renormalisation scheme''
in which superconformality properties were manifest for the whole class of 
superconformal theories; akin to the ``NSVZ'' scheme\cite{JJN} in which the 
gauge $\beta$ function for an $\Ncal=1$ supersymmetric theory in four 
dimensions adopts the simple NSVZ form\cite{NSVZ}. 
Unfortunately it is not clear 
without further calculation whether the present discussion can be extended
to include the presence of flavour fields, in particular whether a redefinition
such as Eq.~(\ref{geny}) could simultaneously remove residual $O(k^4)$ terms 
from the four-loop anomalous dimensions for both bifundamental and flavour 
matter; and while it is tempting to speculate
that indeed $b=\frac43$ (and $a=0$, which simplifies Eq.~(\ref{rat}) in a 
similar way), this must remain a hypothesis for the moment.

\bigskip
\section{Conclusions}
We have shown that on the one hand, superconformal invariance
of Chern-Simons theories in general requires transcendental corrections beyond 
leading order, except for special cases such as $\Ncal=3$ supersymmetry;
and on the other hand, that at leading order 
(and likely beyond) in $N$, $M$, no rational corrections 
are required for a wide class of theories (Class I in our terminology).
Our conclusions could be extended beyond leading order in 
$N$ by the computation of the non-planar diagram in Fig.~\ref{diagsa}(r); which
would confirm or disprove our speculation that $a=0$ and $b=\frac43$ and
hence $G_r=\frac43\pi^2T_4$ (with
$T'_4$ replacing $T_4$ in the Class II case with multitrace deformations). 
If this 
speculation is correct, then at least for the theories considered, and possibly 
more generally, we can 
restore superconformal invariance at four-loop order by a combination 
of transformations of the form Eq.~(\ref{genyy}) (in which we simply have to
substitute the particular form of the superpotential) and Eq.~(\ref{geny}). 
In the latter equation, $\lambda$
will be determined purely by the field content and could be specified by
extending our calculation to the case of non-zero $h_{1-4}$ so that 
we could use the exact superconformality of the $\Ncal=3$ theories to determine 
$R$ in Eq.~(\ref{Zdef}). It woud be interesting to attempt to extend all our 
calculations to the case with flavour matter, especially to see if general
expressions can still be given for the coupling redefinitions required.

\bigskip
 
\bigskip

\noindent
{\large{\bf Acknowledgements\\}}
One of us (CL) was supported by a University of Liverpool studentship. IJ is 
grateful for useful discussions with John Gracey and Tim Jones. We thank the 
referee for drawing our attention to Ref.~\cite{GY}.

\bigskip
\bigskip

\noindent

{\large{\bf{Appendix}\hfil}}

In this appendix we list our superspace and supersymmetry conventions,
which follow those of Ref.~\cite{penatib}.
We use a metric signature $(+--)$ so that a possible choice of $\gamma$ matrices
is $\gamma^0=\sigma_2$, $\gamma^1=i\sigma_3$, $\gamma^2=i\sigma_1$
with
\be
(\gamma^{\mu})_{\alpha}{}^{\beta}=(\sigma_2)_{\alpha}{}^{\beta}, 
\ee
etc. We then have 
\be
\gamma^{\mu}\gamma^{\nu}=\eta^{\mu\nu}-i\epsilon^{\mu\nu\rho}\gamma_{\rho}.
\ee
We have\cite{penatib}
 two complex two-spinors $\theta^{\alpha}$ and $\theta^{\alpha}$ 
with indices raised and lowered according to
\be
\theta^{\alpha}=C^{\alpha\beta}\theta_{\beta},\quad
\theta_{\alpha}=\theta^{\beta}C_{\beta\alpha},
\ee
with $C^{12}=-C_{12}=i$. We then have
\be
\theta_{\alpha}\theta_{\beta}=C_{\beta\alpha}\theta^2,\quad
\theta^{\alpha}\theta^{\beta}=C^{\beta\alpha}\theta^2,
\ee
where 
\be
\theta^2=\frac12\theta^{\alpha}\theta_{\alpha}.
\ee
The supercovariant derivatives are defined by
\bea
D_{\alpha}=&\pa_{\alpha}+\frac{i}{2}\thetabar^{\beta}\pa_{\alpha\beta},\\
\Dbar_{\alpha}=&\pabar_{\alpha}+\frac{i}{2}\theta^{\beta}\pa_{\alpha\beta},
\eea
where
\be
\pa_{\alpha\beta}=\pa_{\mu}(\gamma^{\mu})_{\alpha\beta},
\label{pdef}
\ee
satisfying
\be
\{D_{\alpha},\Dbar_{\beta}\}=i\pa_{\alpha\beta}.
\ee
We also define
\be
d^2\theta=\frac12d\theta^{\alpha}d\theta_{\alpha}
\quad d^2\thetabar=\frac12d\thetabar^{\alpha}d\thetabar_{\alpha},
d^4\theta=d^2\theta d^2\thetabar,
\ee
so that
\be
\int d^2\theta\theta^2=\int d^2\thetabar\thetabar^2=-1.
\ee
The vector superfield $V(x,\theta,\thetabar)$ is expanded in Wess-Zumino gauge
as
\be
V=i\theta^{\alpha}\thetabar_{\alpha}\sigma+\theta^{\alpha}\thetabar^{\beta}
A_{\alpha\beta}-\theta^2\thetabar^{\alpha}\lambdabar_{\alpha}
-\thetabar^{2}\theta^{\alpha}\lambda_{\alpha}+\theta^2\thetabar^{2}D,
\ee
and the chiral field is expanded as
\be
\Phi=\phi(y)+\theta^{\alpha}\psi_{\alpha}(y)-\theta^2F(y),
\ee
where
\be
y^{\mu}=x^{\mu}+i\theta\gamma^{\mu}\thetabar.
\ee

In the main text, our results were given in terms of a basis of momentum
integrals. The results for their divergences were computed in 
Ref.~\cite{MSS,MSSa}
and are listed below
\bea
I_4&=&
-\frac{1}{2\epsilon^2}+\frac{2}{\epsilon}\nn
I_{22}&=&
-\frac{1}{\epsilon^2}\nn
I_{4bbb}&=&\frac{\pi^2}{2\epsilon}\nn
I_{42bbc}&=&
\frac{2}{\epsilon}\nn
I_{422qAbBd}&=&\frac{1}{4\epsilon^2}+\frac{1}{\epsilon}
\left(\frac54-\frac{\pi^2}{12}\right),\nn
I_5&=&\frac14I_4-\frac58I_{22}-I_{4bbb}+I_{42bbc}-2I_{422qAbBd}
=-\frac{\pi^2}{3\epsilon}.
\label{momint}
\eea
Note that our definitions of $I_4$ etc differ by a factor of $F^{-4}$ from those
of Ref.~\cite{MSS}.

\begin{figure}
\includegraphics{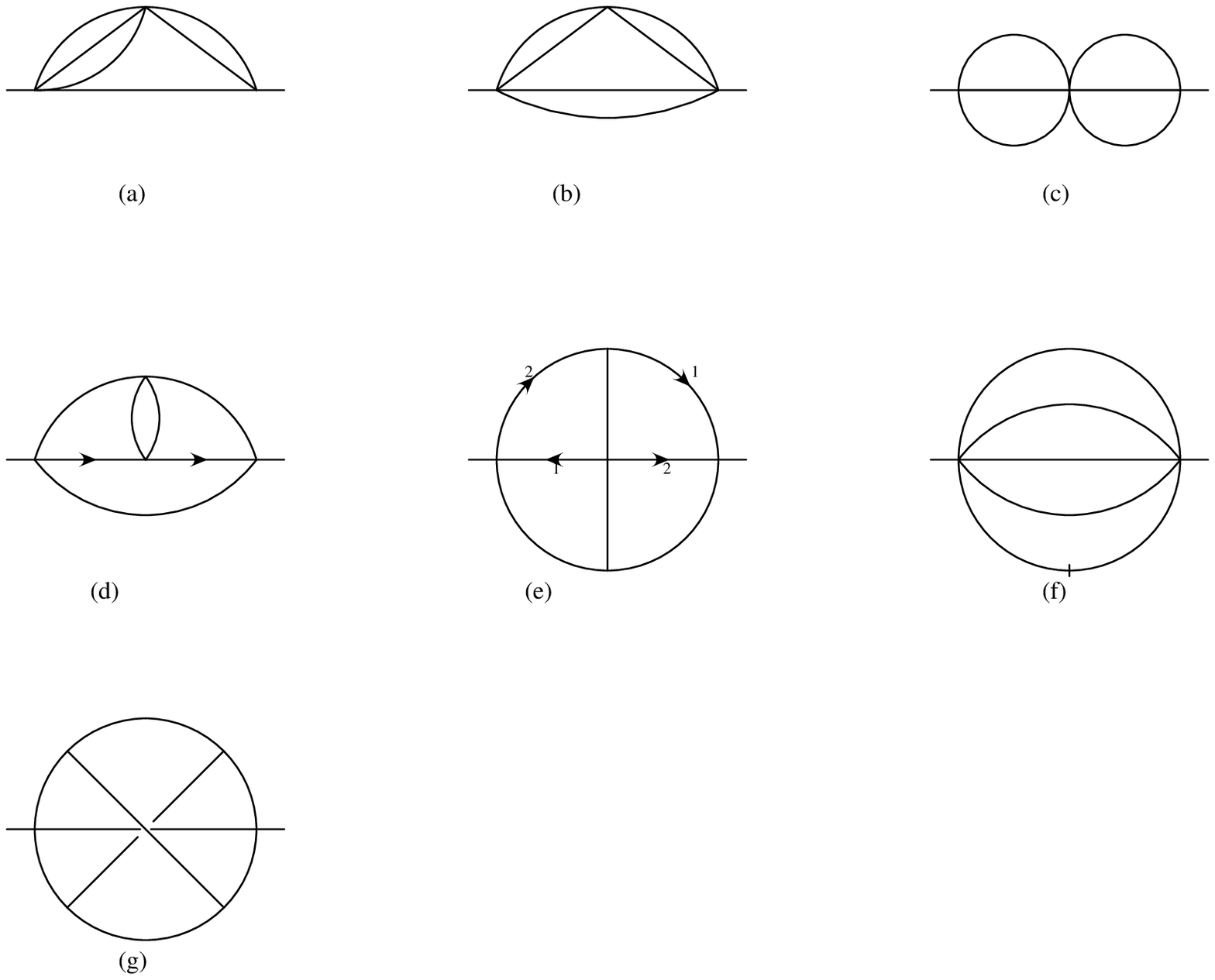}
\caption{Momentum integrals}
\label{allmom}
\end{figure}

Figs.~\ref{allmom}(a)-(f) depict $I_4$, $I_{4bbb}$,
$I_{22}$, $I_{42bbc}$, $I_{422qAbBd}$  and $J_4$ respectively. In the momentum
integral for the so far uncomputed Fig.~\ref{allmom}(g), there is a trace over a series of 
$p^{\mu}\gamma_{\mu}$ in order around the perimeter.

\end{document}